\documentstyle[12pt]{article}

\begin{document}
\vspace*{.3cm}
\begin{flushright}
\large{CINVESTAV-FIS-15-94}
\end{flushright}
\vspace{1.0cm}
\begin{center}
\LARGE{\bf Tensor interactions and $\tau$ decays}
\end{center}
\vspace{.8cm}
\begin{center}\Large
J. J. Godina Nava and G. L\'opez Castro\\
\vspace*{.4cm}
{\normalsize{\it Departamento de F\'\i sica, Cinvestav del IPN,\\
\it Apdo. Postal 14-740, 07000 M\'exico, DF. MEXICO}}
\vspace{.4cm}
\end{center}
\thispagestyle{empty}
\vspace{.4cm}
\centerline{ \bf Abstract}
\vspace{.5cm}

  We study the effects of charged tensor weak currents on the
strangeness-changing decays of the $\tau$ lepton.
   First, we use the available information  on the $K^+_{e3}$ form
factors to obtain BR$(\tau^- \rightarrow K^-\pi^0 \nu_{\tau})\sim
{\cal O}(10^{-4})$ when the $K\pi$ system is produced in an antisymmetric
tensor configuration. Then, we propose a mechanism for the
direct production of the $K_2^*(1430)$ in $\tau$ decays. Using the
current upper limit on this decay we set a bound on the symmetric tensor
interactions.

\vspace{.7cm}
\vspace{5cm}

\newpage
\setcounter{page}{1}
\vspace{2cm}

\begin{center}
\bf 1. Introduction.
\end{center}

   The $\tau$ lepton is the only charged lepton massive enough to decay
into hadrons. This property serves to test interesting Standard Model
(SM) predictions in a clean way. In particular, one can also study several
properties of the charged vector ($\rho(770),\ K^*(892)$) or axial
($a_1(1260)$) mesons produced in $\tau$ decays
 because their production mechanism is free from strong interactions
complications.

   Experimentally, the production mechanism for tensor mesons is of
hadronic origin. For example,
the $a_2(1320)$ is observed in $\pi p$ collisions or in $J/\Psi$ decays
while the $K_2^*(1430)$ is produced in $Kp$ experiments [1]. These
tensor mesons can not be produced by a leptonic mechanism because of
the V or A character of the electromagnetic and weak interactions in the
Standard Model.

  The aim of this letter is to estimate
the effects of tensor interactions in strangeness-changing decays of
the $\tau$ lepton. For the purposes of this paper, it is convenient to
start by introducing some terminology.
We will call an
{\em antisymmetric} tensor interaction to the low energy effective
Lagrangian which involves the product of antisymmetric fermionic currents of
the form
$J_{[\mu\nu]} \sim \overline{\psi} \sigma_{\mu\nu} \psi'$, while the {\em
symmetric} tensor interaction will involve the product of the currents
$J_{\{ \mu\nu\}}
\sim \overline{\psi}\Sigma_{\mu\nu} \psi'$, where $\Sigma_{\mu\nu}$ is a
symmetric tensor involving Dirac gamma matrices.

    Let first argue that on-shell tensor particles ($J^P=2^+$) can not be
produced by the SM interactions. The $V-A$ structure of the
weak charged currents at tree level
does not allow the production of tensor mesons in $\tau$ decays (the
hadronic matrix element $<T|\overline{ q} \gamma_{\mu} (1-\gamma_5)u |0>$,
$q=d,s$ vanish identically).
At the one-loop level, a tensor vertex of the form $\sigma_{\mu\nu} q^{\nu}$
($q^{\nu}$ the four-momentum transfer) can be induced in
the SM by the first order QCD corrections to the vertex
$\overline{q} q'W^{\pm}$ [2].
However, the orthogonality conditions on the polarization tensor
$\varepsilon_{\{ \mu\nu\}}$ describing the tensor particle, forbid the
production of this particle when it is on-shell (the only antisymmetric
tensor that can be written out of $q^{\mu}$ and
$\varepsilon_{\{\mu\nu\} }$ is $<T |
\overline{q} \sigma_{\mu\nu}  u |0> =
\epsilon_{\mu\nu\alpha\lambda}q_{\beta}(\varepsilon^{\{\alpha\beta\}}
q^{\lambda}-\varepsilon^{\{ \lambda \beta\} }q^{\alpha})$ which vanishes
because $q_{\mu}\varepsilon^{\{ \mu\nu \} }=0$). As an alternative, in
this paper we will consider the $|\Delta S|=1$ flavor partner of the
energy-momentum tensor as a possible mechanism for the production of the
$K_2^* (1430)$  meson in $\tau$ decays. Note that symmetric tensor
interactions cannot be generated from radiative corrections to the $V-A$
vertices [2].

 The search for tensor currents dates from the beginning of the weak
interaction theory. Recently, the existence of tensor fermionic
interactions has been raised in several contexts.  For instance, the
presence of tensor antisymmetric interactions has
been suggested in order to explain the apparent problems
observed in $(a)$ the $\pi \rightarrow e\nu \gamma$ decay rate [3] and,
$(b)$ the
measurement of a non-zero tensor term in the decay $K^+ \rightarrow \pi^0
e^+ \nu_e$ [4] (see also Ref. [5]). One also observes the presence of
tensor fermionic currents in the context of the effective
Lagrangian formulation for the low energy weak interactions [6].
Since the tensor interactions in $K^+_{e3}$ decays is
closely related to this work, let us first discuss it in more detail.

  When the V--A requirement for the weak interactions is relaxed, the
decay amplitude for the $K\rightarrow \pi l^+ \nu_l$ ($K_{l3}$) decays
can be written as follows (Ref. [1], p. 1530-1531):
\vspace{.6cm}
 \begin{eqnarray}
{\cal M} &\propto f_+(q^2) [(P_K + P_{\pi})_{\mu} \overline{l} \gamma_{\mu}
(1+\gamma_5) \nu ] + f_-(q^2)[m_l \overline{l} (1+\gamma_5) \nu ]
\nonumber \\  & \nonumber \\ & \ \ + 2m_Kf_S \overline{l} (1+\gamma_5) \nu +
\frac{\textstyle2f_T}{\textstyle m_K} (P_K)_\lambda (P_{\pi})_{\mu}
\overline{l} \sigma_{\lambda\mu} (1+\gamma_5) \nu,
\vspace{.7cm}
\end{eqnarray}
where $q^2=(P_K-P_{\pi})^2$.
 The form factors $f_+,\ f_-$ are associated to the vector
hadronic current of the SM and, in the SU(3) limit, they are normalized
such that $f_+(0)= (1,\ 1/\sqrt{2})$ and $f_-(0)=(0,\ 0)$ for
$(K^0_{l3},\ K^+_{l3})$,  respectively [7]. $f_S,\ f_T$ are
{\em
scalar} and {\em tensor} form factors; their non-zero values would
indicate signals for physics beyond the standard model.

Observe that the last term in Eq. (1) is just a convenient
parametrization introduced to analyse the experiment. This amplitude
could be associated, for example, to an effective Lagrangian that couples
two antisymmetric currents,
\vspace{.6cm}
\[
{\cal L} = -\ \sqrt{2}G_F \tilde{f}_t \overline{u}\sigma_{\alpha\lambda}
d' \left( \frac{q^{\alpha}q^{\beta}}{q^2} \right) \overline{l}_R
\sigma_{\beta\lambda} \nu_L
\vspace{.6cm}
\]
as proposed by Chizhov in Ref. [5] (see the Appendix). In this case, the
last
term of Eq. (1) would arise from the following parametrization of the
hadronic matrix element: \vspace{.6cm}
\[
<\pi | \overline{u} \sigma_{\mu\nu} s| K> \sim
(P_K)_{\mu}(P_{\pi})_{\nu} - (P_{\pi})_{\mu} (P_K)_{\nu} .
\vspace{.6cm}
\]

In the case of $K_{e3}$
decays, the observables are not sensitive to $f_-$ and this allows in
principle to study the effects of $f_S$ and $f_T$. Surprisingly, the
experimental results reported in [1] indicates $ |f_T/f_+(0)| = 0.38 \pm
0.11$ or equivalently,
\vspace{.6cm}
\begin{equation}
f_T \equiv f_T(0) = 0.27 \pm 0.08
\vspace{.6cm}
\end{equation}
 for the $K^+_{e3}$ decay, which is more than three standard
deviations above zero [8]. In passing, let us mention that some other
discrepancies between theory and experiment are observed in $K$
semileptonic decays, namely the isospin breaking in the ratio $f_+(0,\
K^+_{e3})/f_+(0,\ K^0_{e3})$ and the isospin breaking in the slopes of
the scalar form factors of $K^0_{\mu 3}$ and $K^+_{\mu 3}$ [9].

   In order to clarify this possible experimental evidence for scalar
or tensor antisymmetric
interactions, it would be interesting to have new measurements of the form
factors $f_S$ and $f_T$ at the $\phi$ factory, where one expects
the production of about $10^{10}$ pairs of $K^+K^-$/year [10]. In the
following we first will provide an estimate for the SM contribution to
the $\tau^- \rightarrow K\pi \nu_{\tau}$. In section 3 we assume the
existence of the tensor antisymmetric interactions and consider
its effects in $\tau$ decays. In section 4 we will use the current upper
limit on $\tau^- \rightarrow K_2^{*-} \nu_{\tau}$ to set a bound on
symmetric tensor interactions.

\

\begin{center}
\bf 2. SM contribution to $\tau^- \rightarrow K\pi \nu_{\tau}$.
\end{center}

   The SM contribution to the $\tau \rightarrow K\pi \nu_{\tau}$ decay is
given by the following amplitude:
\vspace{.6cm}
\begin{equation}
{\cal M}_{SM} = \frac{G_F V_{us}}{\sqrt{2}}
\overline{\nu}\gamma^{\mu}(1-\gamma_5) \tau <K\pi | \bar{u}\gamma_{\mu}
s | 0>
\vspace{.6cm}
\end{equation}
where $G_F$ denotes the Fermi constant and $V_{us}$ the relevant
Kobayashi-Maskawa matrix element.

  The hadronic matrix element above can be written as:
\vspace{.6cm}
\begin{eqnarray}
<K(k)\pi(k')| \bar{u} \gamma_{\mu} s | 0> &=& f_+(q^2)(k-k')_{\mu} +
f_-(q^2)q_{\mu} \nonumber \\
&=&f_+(q^2)[(k-k')_{\mu} - \frac{\Delta^2}{q^2}q_{\mu}] +
\frac{\Delta^2}{q^2}f_0(q^2)q_{\mu}
\vspace{.6cm}
\end{eqnarray}
where $q=k+k'$ is the momentum transfer to the hadronic system,
$\Delta^2\equiv m_K^2-m_{\pi}^2$ and $f_+,\ f_0$ are form factors
associated to the $J^P = 1^-,\ 0^+$ configuration of $K\pi$.
 Unlike $K_{e3}$ decays where $q^2 \leq (m_K-m_{\pi})^2$, in
the $\tau$ decay under consideration $(m_K+m_{\pi})^2 \leq q^2 \leq
m_{\tau}^2$. This allows
the possibility to produce the $K\pi$ system in a resonant way:
for example, the $K^*(892),\ K^*_0(1430)$ and the $K^*_2(1430)$ in the
$J^P=1^-,\ 0^+$ and $2^+$ channels, respectively.

   The decay rate corresponding to Eqs. (3,4) is given by:
\vspace{.6cm}
\begin{equation}
\Gamma_{SM}(\tau \rightarrow K\pi\nu_{\tau})=\frac{G_F^2
m_{\tau}^5}{768\pi^3} |V_{us}|^2 I_{SM}
\end{equation}
where
\begin{eqnarray}
I_{SM} =\frac{1}{m_{\tau}^8}\int \frac{dq^2}{q^6} (m_{\tau}^2-q^2)^2
& &\hspace{-1.0cm} \left\{
|f_+|^2(m_{\tau}^2+2q^2)\lambda^{3/2}(q^2,m_K^2,m_{\pi}^2) \right.
 \nonumber \\
&+& \left. 3|f_0|^2 \Delta^4 m_{\tau}^2 \lambda^{1/2}(q^2, m_K^2, m_{\pi}^2)
\right\}
\vspace{.6cm}
\end{eqnarray}
and $\lambda (x,y,z)=x^2+y^2+z^2-2xy-2xz-2yz$.

   We can estimate the decay rates by assuming a
simple Breit-Wigner\footnote{A Breit-Wigner with an energy-dependent
width can be chosen as well.} for the form factors in Eq. (6), namely \[
f_i(q^2)=\frac{f_i(0) m_*^2}{m_*^2-q^2-im_*\Gamma_*}\ \ \ \ \ i=+,\ 0
\]
where $m_*,\ \Gamma_*$ are the resonant parameters of the $K^*(892)$ or
$K^*_0(1430)$ when $i=+$ or $0$, respectively. The form factors at
$q^2=0$ are taken from Ref. [11] to be: $f_+(0)=f_0(0)=0.961\pm
0.008$ and $(0.982 \pm 0.008)/\sqrt{2}$ for the $\overline{K}^0\pi^-$ and
$K^-\pi^0$ cases, respectively.

Using $V_{us}$ and the $\tau$ lifetime given in [1] we
obtain,
\begin{eqnarray}
B(\tau^- \rightarrow K^- \pi^0 \nu_{\tau})&=&(3.35\pm 0.11) \times 10^{-3} \\
B(\tau^- \rightarrow \overline{K}^0 \pi^- \nu_{\tau})&=&(6.18\pm 0.21)
\times 10^{-3} \end{eqnarray}
where the quoted errors arise from the uncertainties in
$m_{\tau},\ \tau_{\tau},\ V_{us}$ and $f_i(0)$.
Adding both results we obtain
$B(\tau^- \rightarrow (K\pi)^- \nu_{\tau}) = (9.53 \pm 0.25)\times 10^{-3}$,
which compares
reasonably well with the experimental value $B(\tau^- \rightarrow
K^{*-}(892) \nu_{\tau})= (1.33\pm 0.09)\%$ [12]. The  numerical
discrepancy between both results, if real,  should be attributed to the
simple Breit-Wigner used to parametrize the form factors.
Finally, the $q^2$-dependence of
$f_0$ is not important because it contributes only 3 \% to Eq. (6).

\

\begin{center}
{\bf 3. Antisymmetric tensor interactions.}
\end{center}

Let us now
consider the antisymmetric tensor contribution to the decay $\tau^-
\rightarrow K^- \pi^0
\nu_{\tau}$. If we assume $e-\tau$ universality  for tensor interactions, we
can write the following decay amplitude for the tensor contribution to
this decay:
\vspace{.7cm}
\begin{equation}
{\cal M} = \frac{G_FV_{us}}{\sqrt{2}} \frac{2f_T(q^2)}{m_K} k_{\lambda}
k'_{\mu} \overline{\nu} \sigma^{\lambda \mu}(1+\gamma_5) \tau,
\vspace{.7cm}
\end{equation}
where $f_T$ is the $q^2$-dependent tensor form
factor.

   The decay rate corresponding to Eq.(9) can be written in the following
form (one can easily check that the tensor amplitude do not interfere
with $f_+,\ f_0$ in the decay rate):
 \vspace{.7cm}
\begin{equation}
\Gamma(\tau \rightarrow K \pi \nu_{\tau}) = \frac{G_F^2 |V_{us}
|^2|f_T(0)|^2}{768 \pi^3m_{\tau}^3m_K^2}I_{AS},
\vspace{.7cm}
\end{equation}
where the integral $I_{AS}$ is given by:
\vspace{.7cm}
\begin{equation}
I_{AS}= \int dq^2(m_{\tau}^2 -q^2)^2 (2m_{\tau}^2+q^2)
\lambda^{3/2}(q^2,m_K^2,m_{\pi}^2)\left| \frac{f_T(q^2)}{f_T(0)} \right|^2
\vspace{.7cm}
\end{equation}
  Notice that the $q^2$ distribution, Eq. (11),  contains a factor
$(2m_{\tau}^2 +
q^2)$ instead of $(m_{\tau}^2 + 2q^2)$ obtained for the $1^-$ channel,
Eq. (6). This could help to
isolate the tensor contribution in $\tau \rightarrow K\pi \nu_{\tau}$ and
have an independent measurement of $f_T$.

 If we set $f_T$ to a constant (see the following paragraph for the
possibility of a
$q^2$-dependent form factor), given in Eq. (2), and compute the branching
ratio corresponding to Eq. (10), we
obtain:
\vspace{.7cm}
\begin{equation}
BR(\tau \rightarrow K^-\pi^0 \nu_{\tau})=(1.8 \pm 1.0) \times 10^{-4}.
\vspace{.7cm}
\end{equation}
which lies one order of magnitude below the SM contribution, Eq. (7).
Thus, if present, it seems possible to achieve a measurement of the
non-resonant
production of $K\pi$ in the antisymmetric tensor configuration in a high
statistic experiment such as a $\tau-charm$ Factory.

  The decay rate given in Eq. (10) would receive an enhancement if the
$K\pi$ system were produced in a resonant way. An exotic candidate for
this resonance would be, for example, the strange hybrid-meson of the
$q\overline{q}g\ (J^{P} = 1^+)$ family (this exotic meson can be described
by an antisymmetric tensor $\varepsilon_{[\mu\nu]})$ [13]. However, this
contribution is inhibited because the hybrid mesons decay preferentially
to final states containing excited $q\overline{q}$ mesons [14].

\

\begin{center}
{\bf 4. Symmetric tensor interactions.}
\end{center}

  We propose that the current$\times$current form of the symmetric tensor
interaction Lagrangian for the strangeness-changing $\tau$ decays is
given by
\vspace{.7cm}
\begin{equation}
{\cal L}^{\Delta S=1} = \frac{G_F}{\sqrt{2}}V_{us} g_t (\overline{\nu}
\Sigma_{\mu\nu} l) (\overline{u} \Sigma^{\mu\nu} s)
\vspace{.7cm}
\end{equation}
where the symmetric tensor $\Sigma_{\mu\nu} \equiv
i(\gamma_{\mu}\buildrel\leftrightarrow\over{\partial}_{\nu} +
\gamma_{\nu}\buildrel\leftrightarrow\over{\partial}_{\mu} )$ involves
 first-order derivatives. The dimension of the effective tensor coupling
$g_t$ is ${\rm (mass)}^{-2}$. We have chosen the above Lagrangian in
order to provide a mechanism responsible for the $\tau \rightarrow
K^*_2\nu_{\tau}$ decay. Although this choice does not exclude the
existence of other tensor structures, we have used this Lagrangian for
simplicity. It should be noted that it does not arise from radiative
corrections to vertices with V--A currents [2].

   From the above Lagrangian we get the following amplitude for $\tau(p)
\rightarrow K_2^*(k) \nu_{\tau}(p')$:
\vspace{.7cm}
\begin{equation}
{\cal M} = \frac{G_F}{\sqrt{2}} V_{us} g_t \overline{\nu}
(\gamma_{\mu}Q_{\nu} + \gamma_{\nu}Q_{\mu}) \overline{l} <K_2^* |
\overline{u} \Sigma^{\mu\nu} s |0>
\vspace{.7cm}
\end{equation}
where $Q=p+p'$.
   The hadronic matrix element in the previous equation can be
parametrized as follows:
\vspace{.7cm}
\begin{equation}
<K_2^* | \overline{u} \Sigma^{\mu\nu}s | 0> = g_{K_2^*} m^3
\varepsilon^{\{\mu\nu\}}
\vspace{.7cm}
\end{equation}
where $m$ denotes the $K_2^*$ mass and $\varepsilon^{\{\mu\nu\} }$ its
(symmetric) polarization tensor. With the above definition $g_{K^*_2}$
becomes a dimensionless coupling.

   Let us address a comment on the evaluation of the hadronic matrix
element. Althought it is not a popular idea, it has been suggested in the
literature [15] that the {\em tensor meson dominance} of the
energy-momentum
operator can be assumed in order to give a single parameter description of
 the $\pi\pi$ and $\gamma \gamma$  decays of the $J^P=2^+$ meson $f_2(1270)$.
Since the $K_2^*(1430)$ meson and the $\overline{u} \Sigma_{\mu\nu} s$
operator are flavor
partners of the $f_2(1270)$ and the energy-momentum tensor, respectively,
we can assume the nonet symmetry in order to relate Eq. (15) and the
corresponding
annihilation amplitude of the $f_2$. The use of nonet symmetry gives:
\vspace{.7cm}
\begin{equation}
g_{K_2^*} = \frac{\sqrt{3}}{4}\left(\frac{m_{f_2}}{m} \right)^3 g_{f_2}
\vspace{.7cm}
\end{equation}
where $g_{f_2} = 0.103 \pm 0.011$ has been estimated by Terazawa
[15] by using the $f_2 \rightarrow \pi^+\pi^-$ decay rate.

  The decay rate corresponding to Eqs. (14,15) is:
\vspace{.7cm}
\begin{equation}
\Gamma(\tau \rightarrow K_2^* \nu_{\tau}) = \frac{G_F^2 |V_{us}|^2}{16\pi
M^3} g_t^2 g^2_{K_2^*} m^2 (2M^2 +3m^2)(M^2-m^2)^4
\vspace{.7cm}
\end{equation}
where $M$ denotes the mass of the $\tau$ lepton.

  Finally, if we compare Eqs. (16), (17) and the current upper limit on the
$\tau^- \rightarrow K_2^{*-} \nu_{\tau}$ decay ($\Gamma^{exp} (\tau
\rightarrow
K_2^*(1430)\nu_{\tau}) < 6.7 \times 10^{-12}$  MeV [1]) we obtain the
following bound:
\vspace{.7cm}
\begin{equation}
g_t < 2.6 \times 10^{-6}\ {\rm MeV}^{-2}.
\vspace{.7cm}
\end{equation}

  Eq. (13) will also give a contribution to $\tau^- \rightarrow K^-\pi^0
\nu_{\tau}$. In this case, the hadronic matrix element can be
parametrized as [15]
\begin{equation}
<K\pi | \bar{u}\Sigma_{\mu\nu} s|0>=\frac{g_{K_2^*}g_{K_2^*K\pi} m^2}{m^2
- q^2 -im\Gamma}(k-k')_{\mu}(k-k')_{\nu}
\end{equation}
where $k(k')$ is the momentum of the $K^-(\pi^0)$, and $\Gamma$ is the
total width of the $K_2^*$. The strong coupling
constant $g_{K^*_2K\pi}$ can be determined from $\Gamma^{exp}(K_2^{*-}
\rightarrow K^- \pi^0)=(16.3 \pm 0.6)$ MeV [1] and the expression:
\begin{equation}
\Gamma(K^*_2 \rightarrow K\pi) = \frac{2g_{K^*_2K\pi}^2}{5\pi}\cdot
\frac{|\vec{k}|^5}{m^4}.
\end{equation}
   We can compute the decay rate for $\tau^- \rightarrow K_2^*\nu
\rightarrow K^- \pi^0 \nu$ using the matrix element given in Eq. (19) and
the upper bound given in Eq. (18). We obtain,
\begin{equation}
B(\tau^- \rightarrow (K^-\pi^0)_{symm}\nu_{\tau}) < 9.9 \times 10^{-4}.
\end{equation}
  This upper limit is at the same level as the antisymmetric tensor
contribution given in Eq. (12).

\

\begin{center}
\bf 5. Conclusions.
\end{center}

 We summarize our results. We have studied the effects of tensor
interactions in strangeness-changing $\tau$ decays. Using the information
on the antisymmetric tensor interactions measured in $K^+_{e3}$ decays we
get a
branching fraction for $\tau^- \rightarrow [K^-\pi^0]_{antisym}
\nu_{\tau}$
 which is one order of magnitude below the SM contribution. On the
other hand, we have proposed a mechanism for the direct production of the
$K_2^*(1430)$ in $\tau$ decays. Using the current upper limit on the
$\tau \rightarrow K_2^* \nu_{\tau}$ decay mode we are able to a set bound on
the intensity of the symmetric tensor interactions. Using this upper
bound we have estimated $B(\tau^- \rightarrow
[K^-\pi^0]_{symm}\nu_{\tau}) < 9.9 \times 10^{-4}$.

\

{\bf Acknowledgements}

   The authors acknowledge partial finantial support from CONACyT.

\

\begin{center}
\bf Appendix
\end{center}

  The model of  Chizhov [5], was proposed in order to simultaneously
account for the destructive interference observed in $\pi^+ \rightarrow
e^+ \nu_e \gamma$ [3] and a tensor term in $K^+_{e3}$ reported in [4].

  In Ref. [5], the SM is extended by introducing two Higgs doublets and
two doublets of antisymmetric tensor fields, $T_{\mu\nu}=(T^+_{\mu\nu},
T^0_{\mu\nu})$ and $U_{\mu\nu}=(U^0_{\mu\nu}, U^-_{\mu\nu})$, having
opposite hypercharges ($Y(T)=-Y(U)=+1$) in order to cancel the anomalies.
In this brief summary we use only the interactions of $\nu_e,\ e,\ u,\
d,\ s$ fermions with the tensor fields that are relevant for the
semileptonic processes.

   By assuming quark-lepton universality of the coupling constant $t$,
the SU(2)$_L \times$U(1) invariant interaction gives rise to the
following interaction Lagrangian of the charged tensor fields with
fermions [5]:
\begin{equation}
{\cal L} = \frac{t}{2} \left \{ (\overline{\nu_L} \sigma^{\mu\nu} e_R +
\overline{u}_L \sigma^{\mu\nu} d_R)T^+_{\mu\nu} + \overline{u}_R
\sigma^{\mu\nu} d_L U^+_{\mu\nu} + h.\ c.\right\}
\end{equation}
where $ d,\ u$ are interaction eigenstates. The tensor field $U_{\mu\nu}$
couples only to quarks, because only left-handed neutrinos are present.

  After spontaneous symmetry breaking the charged tensor fields become
mixed and the corresponding matrix of propagators, in the $q^2 \ll m^2,\
M^2$ approximation, is given by [5]
\vspace{.4cm}
\begin{eqnarray}
{\cal P} &=& \left ( \begin{array}{cc}
(T^+T^-)_0 & (T^+U^-)_0 \\
(U^+T^-)_0 & (U^+U^-)_0 \end{array} \right ) \nonumber \\
&=& \frac{2i}{m^2 - M^2} \left( \begin{array}{cc}
\Pi(q) & - {\bf I} \\
- {\bf I} & M^2 \Pi(q)/m^2 \end{array} \right)
\vspace{.6cm}
\end{eqnarray}
where $m,\ M$ are the mass parameters associated with the {\em vev}'s of
the two Higgs doublets, $(XY)_0$ denote the corresponding Green
functions of $X$ and $Y$, and
\vspace{.6cm}
\begin{equation}
\Pi_{\mu\nu\alpha\beta}(q) = {\bf I}_{\mu\nu\alpha\beta} -
\frac{q_{\mu}q_{\alpha}g_{\nu\beta}-q_{\mu}q_{\beta}g_{\nu\alpha}-
q_{\nu}q_{\alpha}g_{\mu\beta} + q_{\nu}q_{\beta}g_{\mu\alpha}}{q^2}
\vspace{.6cm}
\end{equation}
with ${\bf I}_{\mu\nu\alpha\beta} = \frac{1}{2}
(g_{\mu\alpha}g_{\nu\beta} - g_{\mu\beta}g_{\nu\alpha})$.

 After diagonalization of ${\cal P}$ and of the quark mass matrix, Eq. (22)
gives rise to the following four-fermion effective Lagrangian [5]:
\vspace{.6cm}
\begin{equation}
{\cal L} = -\ \sqrt{2}G_F \tilde{f}_t \overline{u}\sigma_{\mu\lambda}
d \left( \frac{q^{\mu}q_{\nu}}{q^2} \right) \overline{l}_R
\sigma^{\nu\lambda} \nu_L
\vspace{.6cm}
\end{equation}
where $G_F\tilde{f}_t/\sqrt{2} = t^2/(M^2-m^2)$ and $d=V_{ud}d' +
V_{us}s'$, with $d',\ s'$ the quark mass eigenstates. Observe that the
hadronic current in Eq. (25) does not include a pseudotensor term; this
will give rise to tensor contributions in $\pi^+ \rightarrow e^+ \nu_e
\gamma$ and $K^+ \rightarrow \pi^0 e^+ \nu_e$, as required by
experiment, but would leave unchanged the $\pi^+ \rightarrow e^+ \nu_e$
decay rate.


\begin{thebibliography}{99}
\bibitem{ref8}
Note also that the scalar form factor is found to be $|f_S/f_+((0)| =
0.084 \pm 0.023$ (see Ref. [1]). Those averages are dominated by a single
high statistic experiment (Ref. [4]). However, the effects of tensor or
scalar interactions have not been observed [1] for the $K_L \rightarrow
\pi^{\pm} e^{\mp} \nu_e$ case.
\end{thebibliography}
\end{document}